# Investigating the opportunities of using mobile learning by young children in Bulgaria

Radoslava Kraleva[#], Aleksandar Stoimenovski[#], Dafina Kostadinova[*], Velin Kralev[#]

[#] Department of Informatics, South West University "Neofit Rilski", Blagoevgrad, Bulgaria

[*] Department of Germanic and Romance Studies, South West University "Neofit Rilski", Blagoevgrad, Bulgaria

*Abstract* – **This paper provides an analysis of literature related to the use of mobile devices in teaching young children. For this purpose, the most popular mobile operating systems in Bulgaria are considered and the functionality of the existing mobile applications with Bulgarian interface is discussed. The results of a survey of parents' views regarding the mobile devices as a learning tool are presented and the ensuing conclusions are provided.**

I. INTRODUCTION

Mobile technologies have significantly changed the life of modern society. Nowadays, most people have not only a phone, but most often they use a smart phone, a tablet and computer as well. These devices are considered as mobile devices[1] and portable devices[2]. Mobile devices most often have their own mobile operating system approaching closer to the operating system[3] desktop and portable computing devices.

These lightweight, comfortable and compact mobile devices are attractive to children. They use them for gaming, watching movies, listening to music, chatting with friends and communicating with the global world. However, all these features are controlled only by a few taps. There is no need to use any additional keyboard or mouse. Children grow up with all these devices around them and acquire their usage before they have started to speak or walk.

The rapid development of hardware and software technologies is a prerequisite for many new ways of using them in modern life [1]. This resulted in adopting a new concept in training called ubiquitous learning (U-learning) which combines traditional classroom training and the possibility of access to educational resources on the Web anytime and anywhere [2], [3], [4]. A number of scientists, psychologists, educators and software developers focused their attention on this new kind of education for children based on applications for mobile and portable devices [5].

On this background the main goal of this study is to research and analyze the usage of mobile devices in the education of young children in Bulgaria. The tasks of this study are:

- To review the most commonly used mobile operating systems in Bulgaria, as well as the possibility of using the existing applications with interface in the Bulgarian language in the training of young children;

- To make an overview of the scientific publications on the topic related to the use of mobile learning;

- To analyze the opinions of parents regarding the use of applications for mobile devices in the education of their children in Bulgaria.

II. APPLICATIONS FOR MOBILE DEVICES WITH BULGARIAN INTERFACE FOR CHILDREN

There are three competing mobile operating systems, Android of Google Inc., Windows Phone and iOS of Microsoft to Apple (Fig. 1) on the Bulgarian markets. Every day users around the world download thousands of apps from the market of these operating systems. Developers rapidly create new and new applications that are added to various markets to meet the needs of consumers.

According to the statistics provided by Microsoft last year [7], there were 669,000 applications in the Windows Store and hundreds of new ones were added every day. This is a small number compared to the markets of Google and Apple.

---

[1] *Mobile devices* – These are all wireless devices with their own operating system, and using Wi-Fi, 3G or 4G internet connection.

[2] *Portable devices* – These are laptops, notebooks, netbooks, ultrabook.

[3] *Mobile operating systems* – Operating systems for mobile devices.

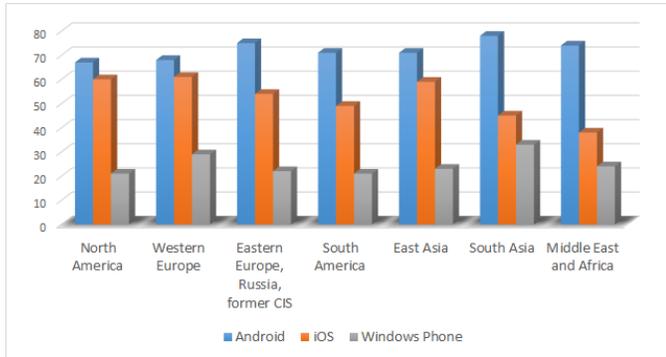

Figure 1. *Selected mobile platforms used by app developers worldwide as of 1st quarter 2014, by region [6]*

The statistics provided by Gartner Inc. [8] reveals that a large proportion of the downloaded applications belongs to the applications for free downloads versus paid applications (Fig. 2).

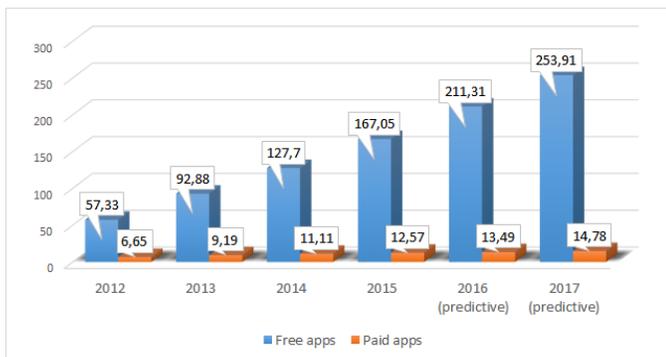

Figure 2. *The number of downloaded free and paid mobile applications for the period 2012 to 2017 (2017 is predictive) according to Gartner [8]*

As it can be seen from the diagram (Fig. 2), most companies and independent developers and users rely on free apps. Often to achieve higher profits other techniques are applied such as adding unwanted ads to free applications, additional modules for which the consumer must pay. In most cases, all applications require a continuous connection to the Internet.

Most downloaded mobile applications refer to games or some other kind of entertainment. Small is the share of software for learning. And even less of it can be used safely by young children, as most of the free applications include pop-up ads or paid modules.

Some of the applications that could be used in teaching young children and are available at Google Play are: "Kids Numbers and Math FREE" (free; studying numbers, counting to 20, addition, subtraction; aged 4 to 6 years); "Number Games: Math for Kids" (Free; Studying objects, shapes, time, numbers, multiplication, division, addition and subtraction; Aged 6 to 10 years); „Baby Puzles" (Free; Recognition of shapes and developing fine motor skills; Aged 0 to 3 years); "LetterSchool free - write abc" (Free; Studying writing uppercase and lowercase letters and numbers up to 10; Aged 0 to 6 years); "Game Kids free 3" (Free; Studying objects, numbers and letters connecting the dots, development of memory and logic; Aged 0 to 6 years) and many others.

Many of these applications are free for download, but the modules providing more features are paid. In the above applications the problem that appears is that there is no support for the Bulgarian language and some games require the assistance of the parent. In other applications, the child is not able to use them at all since a good command of English is needed.

However, applications in the Bulgarian language were also found on Google Play such as „Букви, цифри, цветове Безплатно" /Letters, figures, colors Free/ (Free; learning letters, numbers and colors; Aged 0 to 6 years), „БГ Срички" /BG Syllables/ (Free; Studying writing and pronunciation of Bulgarian words; Aged 2 to 8 years), „Уча АБВ" /Learn ABC/ (Paid; Study the style of writing Bulgarian letters; Aged 2 to 8 years); „БГ Буквар" /BG Primer/ (Free; Learning the letters; Aged 0 to 6 years), and others.

In these applications what is considered as a problem is the presence of unwanted ads and limited functionality compared to alternative applications in English. Many of parents' comments regarding the applications are negative and disapproving, which leads to bad rating of the applications.

On the same date in Apple's App Store a number of applications of the type listed above were found; they had the problem of language and ads. Here applications with support for Bulgarian language that can be used in teaching children at an early age were discovered, too. Such are the „Буквите" /The letters/ (Paid; Pronunciation and spelling of words; Aged 0 to 6 years), "Мозайка" /Mosaic/, "Българските букви и цифри"/The Bulgarian letters and numbers/ (Free; Knowledge of letters, numbers and colors; Aged 0 to 6 years), "Букви с витамини" /Alphabet of vitamins/ (Paid; Pronunciation and writing of letters; Aged 0 to 6 years) and others. No applications associated with the development of mathematical knowledge in children with Bulgarian language interface were found.

Some applications were found in Windows Store which can be used in the training of young children. Such are the "Kids Play & Learn" (Free; Learning the colors, objects, numbers, sounds, time, mathematics, puzzles and languages; Suitable for all ages), "Kids Play Math" (Free; Studying of the numbers and simple math with addition, subtraction, multiplication and division; Aged 0 to 12 years), "Baby Play & Learn" (Paid; Promotes the study of flowers, animals, fruits and objects; Aged 0 to 3 years), "MeSchoolTeacher" (Free; Studying of the numbers, letters and objects; Aged 0 to 17 years), "Kids Preschool Learning Games" (Paid; Studying of the mathematics, numbers, letters, words, objects; Ages 3 to 6 years) and many more. All reviewed applications are supported mainly in English (United States); some of them use several languages, German, Spanish, Russian, but these applications do not support the Bulgarian language.

The wide variety of different and no so much different applications in the stores of Android, Apple and Microsoft lead to the unpleasant trend presented in [9], according to which more and more children in the US spend more time playing on a mobile device, instead of doing some sport or reading books.

As a result of the research presented in this section, it can be concluded that there are still problems related to the quality development of applications for mobile learning. Moreover, in most cases, developers seek quick profits by relying on advertisers or paid modules.

III. TEACHING TO STUDENTS THROUGH MOBILE DEVICES

Over the past few years there has been an increasing interest in the application of non-standard approaches to teaching. Such approaches are based on the use of mobile applications for science learning. Detailed analysis of the design, feasibility and the results achieved by students, when using mobile learning (M-learning,) is presented in [10]. The authors of this paper would recommend more research on mobile learning applications, e.g., using them with more varied science topics and diverse audiences.

In [11] a survey based on more than 110 references in regard to the benefits of using various mobile devices such as laptops, mobile phones and other in the learning process has been done. Based on this, a number of conclusions have been drawn. This paper does not provide any survey of parents of young children who are going to use such devices in their future learning and for various types of entertainment as well.

In [12] the application Mobile Plant Learning Systems is presented; it is used in elementary schools in Taiwan. This software is related to learning in an elementary-school-level botany course, in particular, the classification and detailed information of the different types of plants. The access to the information is through Wi-Fi, and the mobile operating system used is Microsoft Windows Mobile 5.0 Pocket PC Phone Edition. This application is installed on personal digital assistants (PDAs) and the access to the necessary training materials is at any time, regardless of the location of students and teachers. In [12] the effectiveness of the mobile learning is investigated, a survey to obtain feedback from students has been done as well.

Another mobile learning application "ThinknLearn", assisting students in high school to create hypotheses is presented in [13]. Positive results from its practical use in obtaining new knowledge and stimulating deductive thinking in students are observed.

In [14] the opinion of students from preschool to 12th grade (3 to 18 years old) and their parents of the use of mobile devices in the course of their studies is discussed. The data of 2392 parents and 4164 children are processed; children are categorized by gender and age.

In [15] the effects of m-learning, by tracking the development of cognitive skills of young children using the mobile application "Jungle Adventure", for Android and iOS is presented. The working languages of this application are 4 (English, French, German and Spanish). A survey was conducted among 56 children aged 2 to 4 years through the analysis of an independent psychologist. They had to fill in a survey at the beginning of the experiment and every day for three weeks after using the app for about 7.5 minutes. The results showed that about 38% of children have improved cognitive knowledge of colors, letters, objects.

As a result of this study, which cannot be exhaustive enough, it can be concluded that at present very few studies related to young children and their achievements when learning with mobile technologies (M-learning) are available. As a matter of fact, such a study has never been done before in Bulgaria. Therefore, we can say that this is an interesting problem area that is to be analyzed and developed.

IV. RESEARCH AND ANALYSIS OF THE PARENTS' OPINION

Studying the parents' opinion in Bulgaria is important to determine the possibility of implementing mobile learning from early childhood.

Similar analysis of the parents' opinion in the USA, who have at least one child in preschool, is made in [9]. Also, the quality of the education received in the target group, consisting of 90 children, aged 3 to 7 years is evaluated.

From the above mentioned sources one can conclude that the role of parents in the education of children and the successful use of applications for mobile devices in this process are closely related. Therefore, the study of parents' attitudes in Bulgaria on this important issue is of great significance.

To examine the opinion of parents of children under the age of 10 a questionnaire has been developed. It consists of 17 questions: 2 of them are free response, all the others are closed.

The opinion of 50 parents of 64 children, of which 76% have one child, 20% have two children, and the remaining 4% have more than 2 children, is investigated.

The age of their children is shown in Figure 3. As it can be seen from the diagram, most of them are at the age of 4 to 7 years (65%). 16 % of all children are 1 to 2 years of age. There are no children with special educational needs among them.

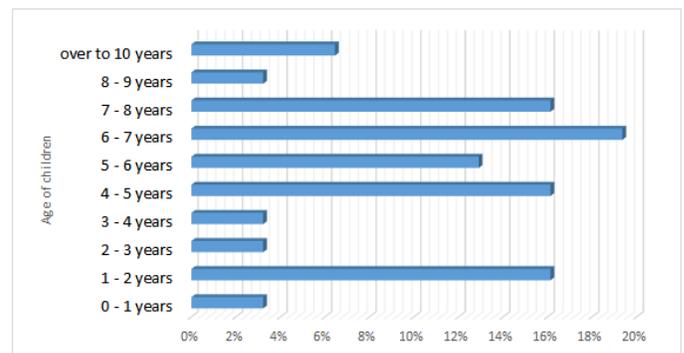

Figure 3. *Age of the children of the surveyed parents*

In Bulgaria some parents of children aged 0 – 2 years prefer to look after them on their own not using the service of the nursery schools; other parents decide that their children would not attend the afternoon classes of the preschool. It should be noted that full-time education in elementary school in Bulgaria is not mandatory. Only 24% of the children of the surveyed parents do not attend any school, because this category includes the children up to age 2 whose parents raise and educate them at home. 24% of children attend school half-day, and 52% - an all-day school.

All interviewed parents confirmed that they devote time for further training of their children at home in order to enhance and consolidate the knowledge instructed at school.

This result points to the fact that parents are an essential part of the learning process of children and the development of any innovative training tools must be presented to a wide range of parents or they can even participate in their preparation and creation.

Like the parents from the surveyed literature from different parts of the world, all parents surveyed in Bulgaria allow their children to use computer devices, including mobile devices, regardless of the age of their children.

The period of time that children can use these devices authorized by the surveyed parents is different (Fig. 4). 64% of parents agree that the optimum time is within two hours per day and 20% of them believe that the best period of time should be reduced to one hour a day.

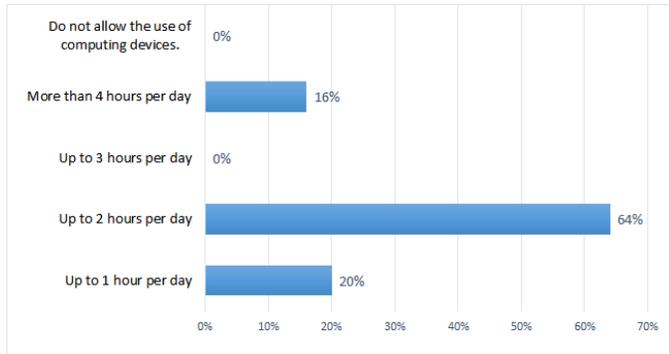

Figure 4. *The period of time during which children can use the computer device*

Most of the families in our survey have different computing devices. This can be seen from the diagram shown in Figure 5 which shows the type of different devices parents allow their children to use. The results here are slightly different from those reported at the beginning of the section.

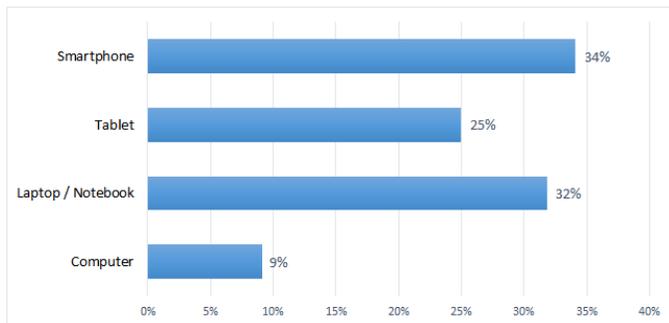

Figure 5. *The types of computing devices that children can use*

This is due to the fact that in Bulgaria, not every child has its own smartphone or tablet, and in most cases he/she uses their parents' devices. The same is true for desktops or portable devices. This is due to the great parental caution related to the safety of children using different computing devices and growing up in good health.

According to the survey of the parents' opinion, 42% of the children use the computer gaming devices, 33% for education and 25% for access to multimedia information such as pictures, music and video (Fig. 6).

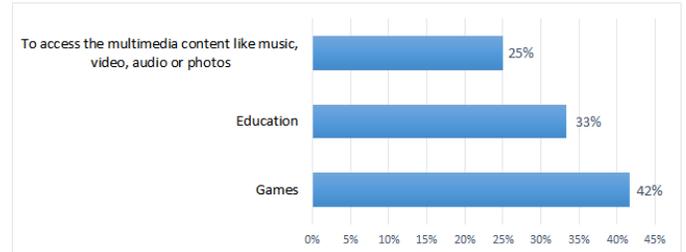

Figure 6. *Activities for which children most often use computing devices*

The operating systems of computing devices that children use are as follows: Android (46%) of Google, Windows (43%) of Microsoft and iOS (11%) of Apple. In this way it is easier to gauge consumer preferences and accordingly, for which operating system an application for mobile learning of young children with interface in Bulgarian to be developed.

The opinion of parents on the applications available for free download in stores on Google, Microsoft and Apple has been investigated. Some parents have responded with more than one answer. The results are rather varied (Fig. 7). The greatest dissatisfaction among the 32% of parents caused the presence of pop-up ads that appear during the use of the application. One solution to this problem is the development of applications without access to the Internet. According to 29% of the parents, another drawback is the lack of support for Bulgarian language in most of the available applications.

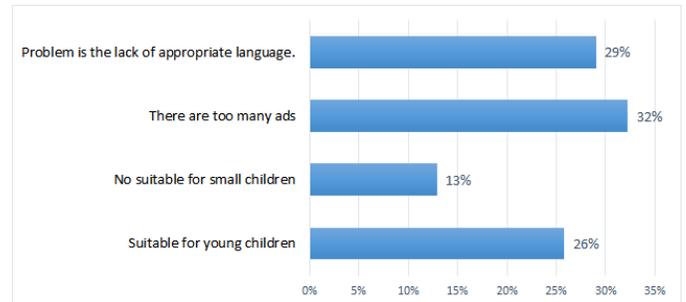

Figure 7. *Opinion on the interface and functionality of applications available for free download in stores on Google Play (Google), Windows Store (Microsoft) and App Store (Apple) for the education of young children in Bulgarian*

An interesting trend in the responses (Figure 6.5) is observed: 26% of the parents believe that there are mobile applications suitable for young children. These parents claim that their children must learn the English language in their early childhood and use it. This is supported by the results obtained here regarding language of the interface their children use. The parents have found that 79% of the applications are in English, 14% in Bulgarian and 7% in other languages.

Furthermore, 92% of all surveyed parents support the idea of using mobile learning in the Bulgarian language, and 64% of them express their positive comments and recommendations. Some of these comments are presented in Table 1.

TABLE I. COMMENTS OF PARENTS ON THE USE OF MOBILE LEARNING IN BULGARIAN

|   | Comments of parents |
|---|---|
| 1 | "It will greatly help in the learning process." |
| 2 | "Great idea!" |
| 3 | "I agree that it is necessary to have such software!" |
| 4 | "The idea is good and it will be something new and interesting for children." |
| 5 | "Modern education needs of such software for mobile learning." |
| 6 | "Such software is absolutely essential!" |

As a result of the present research we may firmly state that 64% of the parents support the education of children by using mobile learning and would be happy to use them in their daily routines. This will ensure complete use of the mobile devices that are up to date and can easily be updated and provide additional knowledge for children. Furthermore, the training will not depend on the location, light, and body position. In this way more freedom of learning is provided and the stress of school and conventional learning can be avoided.

## V. CONCLUSION

In this paper we have reviewed the available applications for mobile devices connected to the educational process of young children. We have classified their functionality and user interface. An analysis of the current literature sources related to the use of mobile devices in teaching young children has been made. A questionnaire survey of parents' opinions concerning the use of applications for mobile phones as a tool for mobile learning has been supplied. The obtained results have been duly presented, analyzed and summarized.

Modern technologies are the tool with the help of which products related to e-learning, M-learning and U-learning are developed. They are a requirement and of great benefit to overcome the barrier of time and space related to knowledge providers and to increase the access to current information. Sometimes the way of providing knowledge depends on the curricula and the different ways of building a mobile learning system. Therefore, when building such systems one should seek to disregard the influence of the technology used on the quality accessible presentation of the targeted educational material. Furthermore, any such system must provide a safe and secure environment for children.